# Hierarchical Sparse Attention Framework for Computationally Efficient Classification of Biological Cells

Elad Yoshai[1], Dana Yagoda-Aharoni[1], Eden Dotan[1] and Natan T. Shaked[1,*]

[1] Department of Biomedical Engineering, Tel Aviv University, Tel Aviv, Israel

[*] Corresponding author: nshaked@tau.ac.il

## Abstract

We present **SparseAttnNet**, a new hierarchical attention-driven framework for efficient image classification that adaptively selects and processes only the most informative pixels from images. Traditional convolutional neural networks typically process the entire images regardless of information density, leading to computational inefficiency and potential focus on irrelevant features. Our approach leverages a dynamic selection mechanism that uses coarse attention distilled by fine multi-head attention from the downstream layers of the model, allowing the model to identify and extract the most salient *k* pixels, where *k* is adaptively learned during training based on loss convergence trends. Once the top-*k* pixels are selected, the model processes only these pixels, embedding them as words in a language model to capture their semantics, followed by multi-head attention to incorporate global context. For biological cell images, we demonstrate that SparseAttnNet can process approximately **15%** of the pixels instead of the full image. Applied to cell classification tasks using white blood cells images from the following modalities: optical path difference (OPD) images from digital holography for stain-free cells, images from motion-sensitive (event) camera from stain-free cells, and brightfield microscopy images of stained cells, For all three imaging modalities, SparseAttnNet achieves competitive accuracy while drastically reducing computational requirements in terms of both parameters and floating-point operations per second, compared to traditional CNNs and Vision Transformers. Since the model focuses on biologically relevant regions, it also offers improved explainability. The adaptive and lightweight nature of SparseAttnNet makes it ideal for deployment in resource-constrained and high-throughput settings, including imaging flow cytometry.

## 1. Introduction

Modern deep-learning approaches to medical image analysis and cell classification typically rely on convolutional neural networks (CNNs) that process entire images pixel by pixel. While being more efficient than fully connected neural networks, these methods are computationally intensive and potentially waste resources by processing uninformative or misleading regions such as background pixels or imaging artifacts. This is particularly problematic in microscopy and cytometry applications, where cells often occupy only a small portion of the field of view, and background areas contain little to no relevant information.

The field of computer vision has seen a paradigm shift with the introduction of attention mechanisms in transformer architectures (Vaswani et al., 2017), which have demonstrated remarkable ability to focus on relevant parts of the input data. However, directly applying these mechanisms to high-resolution microscopy images remains challenging due to the quadratic computational complexity with respect to the number of input tokens.

In this work, we propose SparseAttnNet, a new approach that bridges this gap by dynamically selecting and processing only the most informative *k* pixels from the images using hierarchical attention-driven mechanism. Unlike traditional CNNs that maintain a fixed computational path regardless of input complexity, our method adapts its focus based on the specific characteristics of each image. The key innovation lies in our dynamic adjustment of *k*, the number of pixels processed, which evolves during training based on loss convergence trends.

Flow cytometry, a cornerstone technology in cellular analysis for clinical diagnostics, immunology, and cancer research, stands to benefit significantly from more efficient computational approaches. Current methods for automated cell classification from imaging flow cytometry data often struggle with the variability in cell appearance, imaging artifacts, and the need for real-time analysis in clinical settings (Moen et al., 2019; Perfetto et al., 2004). Our approach addresses these challenges by focusing computational resources on the most discriminative cellular features.

The main contributions of this paper are: (a) A novel dynamic pixels selection mechanism that adaptively finds and selects the optimal number of *k* pixels to process from each image using attention-driven feedback from model loss convergence; (b) A hierarchical attention architecture that first uses a fast, coarse attention module to generate candidate regions and then refines these using a multi-head fine attention mechanism; (c) An integrated fusion scheme where the final prediction is based on both fine attention features and coarse attention features; (d) A distillation loss using Kullback–Leibler (KL) divergence to align the attention distributions across hierarchy levels; (e) A contrastive learning component that enhances feature separation between classes, improving classification robustness.

Our work builds upon recent advances in efficient attention mechanisms (Kitaev et al., 2020; Wang et al., 2020) and point-based neural networks (Charles et al., 2017; Qi et al., 2017), but introduces novel adaptations that leverage sparsity to reduce computational complexity and improve inference speed. The resulting approach not only meets classification performance but also provides interpretable visualizations of the features that drive classification decisions, contributing to the explainability of deep-learning models in specific medical applications.

Cell classification using deep learning has evolved considerably over the past decade. Convolutional neural networks (CNNs) have dominated this field, with architectures like ResNet (He et al., 2015), DenseNet (Huang et al., 2017), EfficientNet (Tan and Le, 2019), and MobileNet (Howard et al., 2019, 2017; Sandler et al., 2018) being adapted for microscopy image analysis. These approaches typically process entire images, regardless of information content distribution. Gupta et al. (Gupta et al., 2019) demonstrated the application of CNNs to flow cytometry data, achieving high accuracy but at significant computational cost. Eulenberg et al. (Eulenberg et al., 2017) applied deep learning to single-cell analysis for detecting cell cycle phases, while Blasi et al. (Blasi et al., 2016) developed CNN-based approaches for white blood cell classification from microscopy images. More recent work has focused on making these models more efficient. Doan et al. (Doan et al., 2020) proposed MicroNet, a lightweight CNN designed specifically for microscopy applications with limited computational resources. However, these approaches still process entire images uniformly, without adapting to the distribution of information within individual samples.

Attention mechanisms have revolutionized computer vision following the success of the Transformer architecture (Vaswani et al., 2017) in natural language processing. Vision Transformer (ViT) (Dosovitskiy et al., 2021) adapted this approach to image classification by dividing images into patches and treating each as a token. While effective, the quadratic complexity of self-attention with respect to the number of tokens makes this approach computationally prohibitive for high-resolution microscopy images.

Several methods have been proposed to address this limitation. Swin Transformer (Liu et al., 2021) introduced a hierarchical architecture with shifted windows to reduce complexity. MViTv2 (Li et al., 2022) incorporated pooling attention to reduce the number of tokens progressively. Most relevant to our work, Meng et al. (Meng et al., 2021) proposed AdaViT, which dynamically adjusts the computation allocated to different parts of the input. These approaches, however, do not specifically address the unique challenges of inputs such as microscopy images, where relevant information is often sparse and unevenly distributed.

Point-based neural networks offer an alternative approach to processing spatially distributed data. PointNet (Charles et al., 2017) pioneered this area by operating directly on point clouds for 3D shape analysis. PointNet++ (Qi et al., 2017) extended this with hierarchical feature learning to capture fine geometric details. While primarily developed for 3D point clouds, these architectures have inspired approaches for processing 2D images as collections of points.

SampleNet (Lang et al., 2020) introduced a differentiable sampling layer that selects representative points from 3D point clouds to improve efficiency while maintaining accuracy. Similarly, Set Transformer (Lee et al., 2019) proposed an efficient attention mechanism for processing sets of points. These methods, however, use a fixed number of points and are not designed to adapt to the varying information content of different samples.

In the domain of sparse image processing, works like submanifold sparse convolution (Graham et al., 2017; Graham and Maaten, 2017) have shown that processing only non-zero pixels can significantly reduce computational requirements. These approaches, however, rely on predefined sparsity patterns rather than learning which pixels are most informative for a specific task.

DynamicViT (Rao et al., 2021) introduced an adaptive token selection mechanism within the Vision Transformer architecture, focusing on pruning tokens to enhance performance in computer vision tasks. Their approach employs an additional classifier to select the most informative tokens at each step, which introduces extra computational complexity and affects the model's efficiency. While their method discards tokens from a fixed grid representation, there is potential to dynamically select the most informative pixels from a sparse representation, allowing more precise focus on relevant image features, especially when the pixels are evenly distributed in the image.

Dynamic neural networks that adapt their computation based on input complexity have gained traction to improve efficiency. Graves (Graves, 2017) introduced adaptive computation time for recurrent neural networks, allowing the model to decide how many computational steps to perform. Building on this concept, MSDNet (Huang et al., 2018) proposed a multi-scale dense network that makes early exit decisions for easy examples. Similarly, SkipNet (Wang et al., 2018) learns to skip layers in deep networks when they are unnecessary for specific inputs.

Our approach bridges these diverse research streams, combining elements of attention mechanisms, pixel-based processing, and dynamic architectures to create a system specifically optimized for efficient and accurate cell classification from microscopy images.

## 2. Methods

### 2.1 Problem Formulation

We address the task of classifying microscopy images of biological cells into predefined categories with minimal computational overhead. Given an input image $I \in \mathbb{R}^{H \times W}$ (where $H$ and $W$ are the height and width in pixels), our goal is to predict the cell class $y \in \{1,2,\ldots,C\}$ where $C$ is the number of classes.

Instead of processing all pixels in the image, we represent the image as a set of pixels, as follows:

$$P = \{p_i = (x_i, y_i, v_i) \mid i \in \{1,2,\ldots,N\}\}, \quad (1)$$

where $x_i$ and $y_i$ are spatial coordinates, $v_i$ is the pixel intensity, and $N = H \times W$ is the total number of pixels. From this set, our model dynamically selects the $k$ most informative pixels, where $k$ is adaptively determined during training, and uses only these pixels to make the classification decision.

### 2.2 Hierarchical Attention Architecture

Our SparseAttnNet model employs a two-stage hierarchical attention architecture to efficiently process microscopy images. The input image first passes through a coarse attention module that rapidly highlights the most important pixels relevant to the classification task. From this attention map, we dynamically sample $k$ pixels with the highest attention weights, where $k$ is a learnable parameter that adapts during training based on loss convergence trend. These selected pixels are enriched and embedded jointly into a high-dimensional space to preserve spatial context before being processed by a multi-head fine attention mechanism. The fine attention calculates the relative importance of each projected pixel and produces weighted feature representations. These features, combined with global context from the coarse module, are fed into the classification branch that produces the final class prediction. The full architecture is described in Fig. 1.

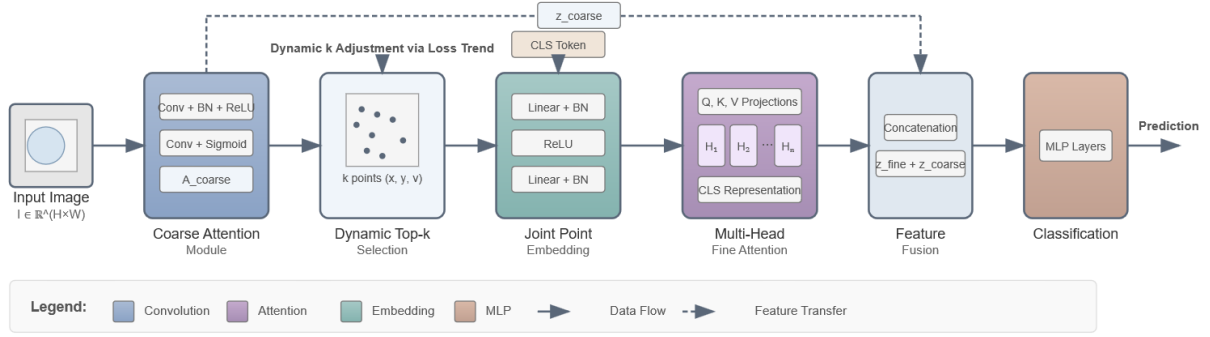

**Figure 1.** Block diagram of the SparseAttnNet architecture. Our architecture first samples the top $k$ pixels based on the coarse attention, embed the selected pixels into higher dimension and applies the fine attention. The output features from the coarse and fine attention modules are then concatenated and proccesed by the classifier to predict the cell type.

### 3.2.1 Coarse Attention Module

The coarse attention module is responsible for quickly inferring candidate regions over the input image $I \in R^{H \times W}$. Its operations are summarized as follows:

**Feature Extraction:** The image is processed by a lightweight convolutional network, as follows:

$$F = f_{\text{conv}}(I), \qquad (2)$$

where $f_{\text{conv}}(\cdot)$ represents a series of convolutional layers with batch normalization and ReLU activations.

**Attention Map Generation:** The feature maps are passed through a sigmoid activation $\sigma$ to generate a coarse attention mask:

$$A_{\text{coarse}} = \sigma(F) \in (0,1)^{H \times W}, \qquad (3)$$

This mask highlights regions that are likely to contain discriminative information.

### 3.2.2 Dynamic Interaction Classifier with Fine Attention

After the coarse attention, the top-$k$ pixels are selected from the image using $A_{\text{coarse}}$. Let $\{(x_i, y_i, v_i)\}_{i=1}^{k}$ denote these pixels, where $(x_i, y_i)$ are the spatial pixel coordinates and $v_i$ is the corresponding pixel intensity.

**Joint Sparse Point Embedding:** Each selected sparse pixel $p_i = (x_i, y_i, v_i)$, where $(x_i, y_i)$ are normalized spatial pixel coordinates and $v_i$ is the intensity, is embedded jointly into a high-dimensional space using a shared embedding function, defined as follows:

$$\mathbf{e}_i = f_{\text{embed}}(x_i, y_i, v_i) \in \mathbb{R}^D, \qquad (4)$$

where $f_{\text{embed}}(\cdot)$ is a neural network trained to map each sparse pixel into a latent representation of dimension $D$. Unlike traditional positional encodings, this approach treats each $(x_i, y_i, v_i)$ triplet as a token (analogous to a word in language models), allowing the network to learn complex interactions between spatial position and intensity.

This formulation offers two major benefits. First, Joint Encoding allows us to learn a unified embedding that captures non-linear dependencies among the three dimensions, instead of linearly projecting position and intensity separately and summing them. Second, the Tokenization of Sparse Pixels treats each selected pixel as a semantically meaningful unit due to the inherent sparsity. The network learns to assign context-aware embeddings, enabling the attention mechanism to operate analogously to transformer-based models that process sequences of discrete tokens.

**Classification (CLS) Token and Global Context**: To enable global reasoning, we append a learnable token $\mathbf{e}_{\text{cls}} \in \mathbb{R}^{1 \times D}$ to the sequence of embeddings, as follows:

$$\mathbf{E} = [\mathbf{e}_1, \mathbf{e}_2, \ldots, \mathbf{e}_k, \mathbf{e}_{\text{cls}}] \in \mathbb{R}^{(k+1) \times D}. \quad (5)$$

After the multi-head attention, the output corresponding to the CLS token is used as the global representation of the image.

**Multi-Head Fine Attention:** The set $\mathbf{E}$ is processed through a multi-head attention mechanism. First, we project the embeddings to obtain values, as follows:

$$\mathbf{V} = \mathbf{E}\mathbf{W}^V. \quad (6)$$

For each head $h = 1, \ldots, H$, we compute:

$$\mathbf{Q}_h = \mathbf{E}\mathbf{W}_h^Q, \quad (7)$$

$$\mathbf{K}_h = \mathbf{E}\mathbf{W}_h^K, \quad (8)$$

where $\mathbf{W}^V$, $\mathbf{W}_h^Q$, and $\mathbf{W}_h^K$ are the projection matrices. We compute attention weights by applying ReLU to the keys and normalizing using efficient normalization (instead of the standard Softmax function):

$$\mathbf{K}_h^+ = \text{ReLU}(\mathbf{K}_h) + \epsilon, \quad (9)$$

$$\mathbf{A}_h = \frac{\mathbf{K}_h^+}{\sum_{j=1}^{k+1} \mathbf{K}_{h,j}^+}, \quad (10)$$

where $\epsilon$ is a small constant for numerical stability. The context and head outputs are computed as follows:

$$\mathbf{C}_h = \mathbf{A}_h^T \mathbf{V}, \quad (11)$$

$$\mathbf{O}_h = \mathbf{Q}_h \mathbf{C}_h. \quad (12)$$

The CLS token representation is then extracted by taking the mean of the outputs at the CLS token position across all heads:

$$z_{\text{fine}} = \mathbf{r}_{\text{cls}} = \frac{1}{H} \sum_{h=1}^{H} \mathbf{O}_{h, k+1}. \quad (13)$$

This $\mathbf{r}_{cls}$ vector serves as our fine-grained global representation $z_{\text{fine}}$ from the attention mechanism.

**Feature Fusion and Final Prediction:** Next, we also extract additional global features from the coarse attention module, denoted by $z_{\text{coarse}}$ (the pooled features from an intermediate layer in the coarse network). These two sources of information are concatenated as follows:

$$f = [z_{\text{fine}}; z_{\text{coarse}}]. \quad (14)$$

Finally, the fused feature vector is passed through a multi-layer perceptron (MLP) with residual connections to obtain the class prediction:

$$y_{\text{pred}} = \text{MLP}(f). \quad (15)$$

This fusion strategy leverages the rapid global context provided by the coarse module along with the refined details captured by the fine attention mechanism.

## 2.3 Loss Formulation

Training the network is guided by a compound loss that enforces accurate classification while ensuring consistency across the hierarchical attention mechanisms.

### 2.3.1 Focal Loss

We supervise our network using focal loss (Lin et al., 2017), which enhances classification accuracy for imbalanced cell datasets by dynamically weighing the loss contribution of each sample as follows:

$$L_{\text{focal}} = -\sum_{i=1}^{N} \alpha_{y_i} \left(1 - p_{y_i}\right)^{\gamma} \log(p_{y_i}), \quad (16)$$

where $p_{y_i}$ is the predicted probability for the correct cell class $y_i$, $\alpha_{y_i}$ is a class-specific weight addressing imbalance (particularly important for rare cell types), and $\gamma$ is the focusing parameter. The modulating factor $\left(1 - p_{y_i}\right)^{\gamma}$ reduces the contribution of easily classified cells and increases focus on challenging cases, such as morphologically similar cell types.

### 2.3.2 Contrastive Loss

To encourage similar samples (intra-class) to have closely aligned representations and dissimilar samples (inter-class) to be further apart, we define a contrastive loss (Khosla et al., 2021):

$$L_{\text{contrast}} = -\frac{1}{B} \sum_{i=1}^{B} \log \frac{\exp(s(z_i, z_i^+)/\tau)}{\sum_{j=1}^{B} \exp(s(z_i, z_j)/\tau)}, \quad (17)$$

where $s(\cdot, \cdot)$ denotes the cosine similarity between normalized embeddings, $\tau$ is a temperature parameter, and $z_i^+$ denotes a positive pair from the same class.

### 2.3.3 Distillation Loss

To improve the coarse attention module and to enforce consistency between the coarse and fine attention modules, we use a KL divergence-based (Kullback and Leibler, 1951) distillation loss

(Hinton et al., 2015). Let $A_{\text{coarse}}$ be the coarse attention map and $A_{\text{fine}}$ be the aggregated attention map obtained from the fine attention scores.

After flattening each map, we normalize them using the Softmax function as follows:

$$P_{\text{coarse}} = \text{Softmax}(A_{\text{coarse}}^{\text{flat}}), P_{\text{fine}} = \text{Softmax}(A_{\text{fine}}^{\text{flat}}). \quad (18)$$

The KL divergence between these distributions provides the distillation loss:

$$L_{\text{distill}} = \text{KL}(P_{\text{coarse}} \parallel P_{\text{fine}}) = \sum_i P_{\text{coarse}}(i) \log \frac{P_{\text{coarse}}(i)}{P_{\text{fine}}(i)}. \quad (19)$$

We freeze the fine attention and apply the gradients to the coarse attention module, so that the learning will go in one direction: from the fine to the coarse attention module.

### 2.3.4 Total Loss

The overall loss is a weighted sum of the focal, contrastive, and distillation losses:

$$L_{\text{total}} = L_{\text{focal}} + \lambda_{\text{contrast}} L_{\text{contrast}} + \lambda_{\text{distill}} L_{\text{distill}}, \quad (20)$$

where $\lambda_{\text{contrast}}$ and $\lambda_{\text{distill}}$ are hyperparameters that balance the contributions of each component.

## 2.4 Dynamic Pixel Selection

A unique feature of SparseAttnNet is its ability to dynamically adjust the number $k$ of pixels used for attention during training. Instead of a fixed $k$, the model adapts based on the loss convergence trend, as follows:

(a) We maintain an exponential moving average (EMA) of the training loss:

$$L_{\text{EMA}} = \beta \cdot L_{\text{EMA}}^{\text{prev}} + (1 - \beta) \cdot L_{\text{current}}. \quad (21)$$

(b) We calculate the trend of the smoothed loss:

$$\Delta L = L_{\text{EMA}} - L_{\text{EMA}}^{\text{prev}}. \quad (22)$$

(c) Based on this trend, we update $k$:

- If $\Delta L \geq 0$ (loss increasing or stagnant): $k \leftarrow \min(k + \Delta k, k_{\max})$.
- If $\Delta L < 0$ (loss decreasing): $k \leftarrow \max(k - \Delta k, k_{\min})$.

(d) We apply momentum to smooth the changes in $k$:

$$k \leftarrow \alpha \cdot k_{\text{prev}} + (1 - \alpha) \cdot k. \quad (23)$$

This dynamic adjustment helps the network achieve an optimal balance between sufficient feature extraction and computational efficiency.

## 2.5 Computational Complexity

Below we analyze the computational complexity of our approach versus Vision Transformers and standard CNNs, to show that our approach achieves significant efficiency gains compared to standard neural network architectures by restricting the computationally intensive attention mechanism to only a subset of image pixels.

**Standard CNNs**: For an image of resolution of $H \times W$ pixels, a convolutional layer with kernel size $K$ operating on $C_{in}$ input channels and producing $C_{out}$ output channels has a computational complexity of approximately of $O(H \cdot W \cdot C_{in} \cdot K^2 \cdot C_{out})$. As image dimensions increase, the computational cost grows quadratically with the number of pixels. For high-resolution microscopy images, this scaling behavior becomes prohibitively expensive, especially when deploying models in resource-constrained environments or for real-time cell classification tasks.

**Vision Transformers (ViTs):** Vision Transformers divide the input image into $N$ non-overlapping patches and process them using self-attention mechanisms. The computational complexity of the standard self-attention operation in ViTs is $O(N^2 \cdot D)$, where $N$ is the number of image patches and $D$ is the embedding dimension. This quadratic scaling with respect to the number of patches makes ViTs computationally expensive for high-resolution images, requiring significant downsampling or aggressive patching strategies that may lose fine cellular details critical for accurate classification.

**SparseAttnNet Pipeline:** Our hierarchical approach strategically distributes computational load. First, the Coarse Attention Module operates on the full image with complexity of $O(H \cdot W \cdot C_{coarse})$, where $C_{coarse}$ is designed to be small through our lightweight network architecture, minimizing the computational burden of processing the entire image. Next, the Multi-Head Fine Attention module operates only on $k$ sparse pixels, yielding a complexity of $O(N_{heads} \cdot k^2 \cdot D)$, where $D$ is the embedding dimension and $N_{heads}$ is the number of attention heads. In our implementation, $k \ll H \cdot W$ and is adaptively reduced during training, making the quadratic term $k^2$ substantially smaller compared to processing the full image. By shifting the bulk of the computational load onto a small, dynamically selected set of pixels, SparseAttnNet achieves a substantial reduction in computational cost while maintaining classification performance. For typical microscopy images in our experiments, where only about 15% of the pixels are processed by the fine attention module, this translates to approximately an 85% reduction in computational requirements compared to standard approaches that process all pixels with equal emphasis.

## 2.6 Implementation Details

Our model is implemented using PyTorch and trained on three types of datasets: images of stained white blood cells (WBCs) from conventional brightfield microscopy, images of WBCs flowing in a microfluidic channel and acquired without staining by a motion-sensitive (event) camera (sparse data), and images of WBCs flowing in a microfluidic channel and acquired without staining using holographic microscopy (highly informative and quantitative optical path difference (OPD) profiles).

**Model Architecture**: Input images are normalized to the range [0, 1]. Positional embeddings use dimension $d = 4$. The number of attention heads is set to $H = 2$. Initial value of $k$ is set to 8000 pixels per image. The minimum $k$ is set to 1500, with a maximum of $H \times W$ (full image). The MLP classifier has a hidden dimension of 512 and includes multiple layers with batch normalization.

**Training Configuration**: We used the AdamW Optimizer with learning rate of 1e-3 and weight decay of 1e-4. Validation split of 20% is used to monitor performance, with data randomly split. We used batch size of 64 for T-cell datasets and 32 for WBC datasets, focal loss of $\gamma = 2.0$, contrastive loss weight of $\lambda_{\text{contrast}} = 0.1$ with temperature $\tau = 0.07$, distillation loss weight $\lambda_{\text{distill}} = 0.02$ with emphasis factor of 2.0, and learning rate schedulers: ReduceLROnPlateau with factor of 0.9 and patience of 5.

**Dynamic $k$ Adaptation**: We used exponential moving average (EMA) with $\beta = 0.2$, momentum for $k$ updates with $\alpha = 0.2$, learning rate for $k$ adjustment of $\text{LR}_k = 80$, and reduction step $\Delta k = 50$.

## 3. Experimental Results

### 3.1 Experimental Datasets

Accurate and efficient cell characterization is essential in biomedical fields as pathology, immunology and disease diagnosis, where rapid analysis and reduced computational costs are critical for practical application. We evaluated SparseAttnNet on three microscopy datasets of cells, each posing distinct classification challenges: conventional WBC images after being stained with Giemsa, cancer cell images during flow acquired by an event camera without staining, and T cells acquired with digital holographic microscopy without staining. In comparison to conventional brightfield imaging, acquisition methods that use motion-sensitive (event) cameras can ease the size of the dataset, acquiring only when the movement appear in the field of view. In addition, quantitative imaging methods, such as digital holography allows stain-free acquisition of the cell dry mass maps, allowing the acquisition of both their structure and content, thus provides a richer imaging data and a better starting point of the classifying network.

*3.1.1 Conventional Brightfield Imaging of Stained WBCs*

Blood analysis is used as the first diagnosis and monitoring tool for many pathological conditions. In the modern clinical lab, flow cytometry (FC) is first used for complete blood count (CBC), for initial inspection, automatically providing a high-throughput counting of the blood-cell types. The present FC machines are not based on imaging, but rather on measurements of integral impendence and light absorption from all cell spatial points. Therefore, there are many ambiguities when classifying the cells. For example, different blood cells or broken cell components might yield the same FC measurements, and thus cell population overlaps occur in the resulting count. Therefore, it is hard to distinguish between certain blood cell pathologies based on FC, for instance, between leukemia and other blood pathologies with increased WBC count. CBC machines in blood diagnostic labs generate warning flags in a large number of borderline cases (e.g., 15-20% of the times in CBCs with 5-class WBC subtyping). Then, optical microscopy review on a stained blood smear is required for visual verification. This latter analysis is typically done manually under a light microscope, after the cells are stained with Giemsa that enhance their imaging contrast and allow analysis of their inner structures. This process is laborious and subjective, providing very low throughput, less than one cell per second. Automatic classifiers can help significantly, especially if the cells are imaged during flow via imaging flow cytometers.

Our WBC image dataset is obtained from Kaggle (Nickparvar, 2022) and consisted of five types of stained WBCs: 1,034 lymphocytes, 234 monocytes, 2,660 neutrophils, 89 basophils, and 322 eosinophils, with 575×575 pixels each. This dataset presents a significant class imbalance challenge, particularly with the limited number of basophil examples (which are typically rare in blood). We resized the images to a dimension of 256×256 and we used 80% of the data for training and 20% for testing. Fig. 2 presents image examples of each cell type.

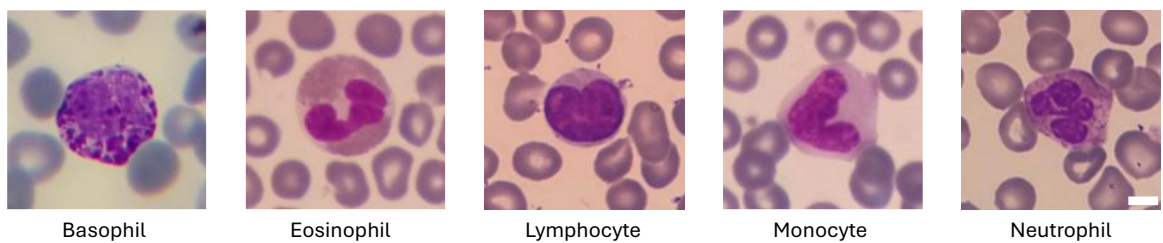

Basophil    Eosinophil    Lymphocyte    Monocyte    Neutrophil

**Figure 2.** Representative examples of stained-based convenentional brightfield microscopy of WBCs. The images showcase the distinct morphological characteristics of lymphocytes, monocytes, neutrophils, basophils, and eosinophils that are essential for accurate classification. The dataset exhibits significant class imbalance, particularly with basophils being the most underrepresented class. The white bar indicates 5 μm.

### 3.1.2 Event Camera Imaging of Cancer Cells

This dataset was acquired in our lab using event-based camera (IDS, UE-39B0XCP-E) under 56.25× magnification, which captures dynamic changes in the scene at the pixel level. The dataset includes grayscale cell images with dimensions of 256×256 pixels (resized to 135×135), containing two classes of colorectal cancer cell lines acquired during flow in a microfluidic channel (imaging flow cytometry): 794 images of SW480 cells (derived from a primary tumor) and 790 images of SW620 cells (derived from a metastatic site of the same patient). These cell lines represent different stages of colon cancer progression of the patient and exhibit subtle morphological differences. The cells were imaged while rolling, thus, the dataset contains multiple projections of the same cell, capturing various orientations and aspects of their morphology. We used 80% of the data for training and 20% for testing. Fig. 3 presents image examples of each cell type acquired by the event camera.

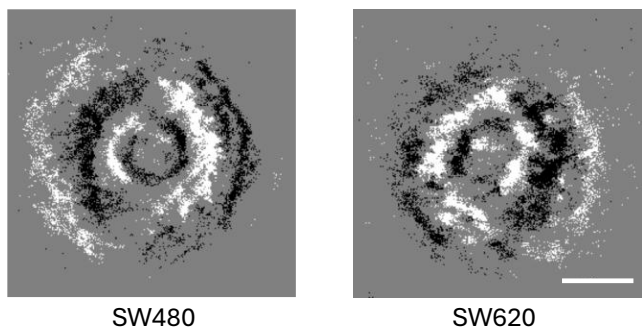

SW480    SW620

**Figure 3.** Two types of cancer cells with different metastatic potential, as acquired by the event camera images during flow. The data appears notably sparse because event cameras only record pixels detecting motion. Furthermore, these cameras only indicate whether a pixel intensity increased or decreased, without capturing all morphological cell values, resulting in limited data representation. The white bar indicates 5 μm.

### 3.1.3 Digital Holographic Microscopy of T-Cells

The dataset includes three classes of T-cells: non-activated T-cells, activated cells after 3 days (ACT) and 434 activated cells after 7 days (ACT7d), with 2,666, 2,234 and 2,434 images, respectively with 135×135 pixels each. This dataset was acquired in our lab using a digital holographic microscopy setup (Cohen et al., 2024). Shortly off-axis holograms are acquired during cell flow in a microfluidic channel without staining the cells. Fig. 4 shows off-axis image hologram examples of each cell type. From the acquired hologram the optical path delay (OPD) profile of the cell is extracted numerically, representing the quantitative dry mass map of the cell. The OPD maps is linked to both the cell morphology and content (via the cell refractive index), enabling the detection of differences between activation states. We used 80% of the data for training and 20% for testing. Fig. 5 presents OPD map examples of each cell type. We trained the network separately on the raw off-axis holograms, which is better for high-throughput processing, since it spared the OPD numerical extraction, and then on the OPD maps, in which the changes between cells are more visual. Thus the first set of the raw hologram, in which the network needs to ignore the off-axis carrier fringe frequency and still detect changes between the cells is very challenging.

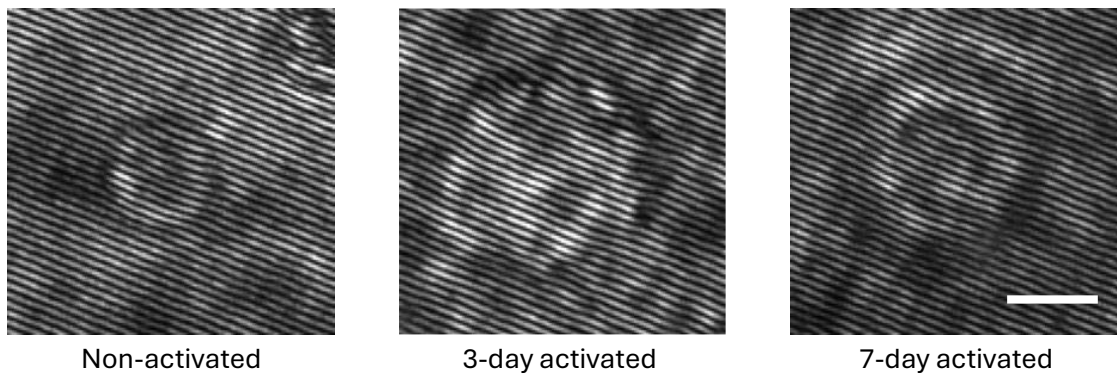

**Figure 4.** Representative cell types from the hologram images dataset. These holographic images highlight the three-dimensional cellular structures, providing additional morphological details that enhance classification accuracy. The white bar indicates 5 μm.

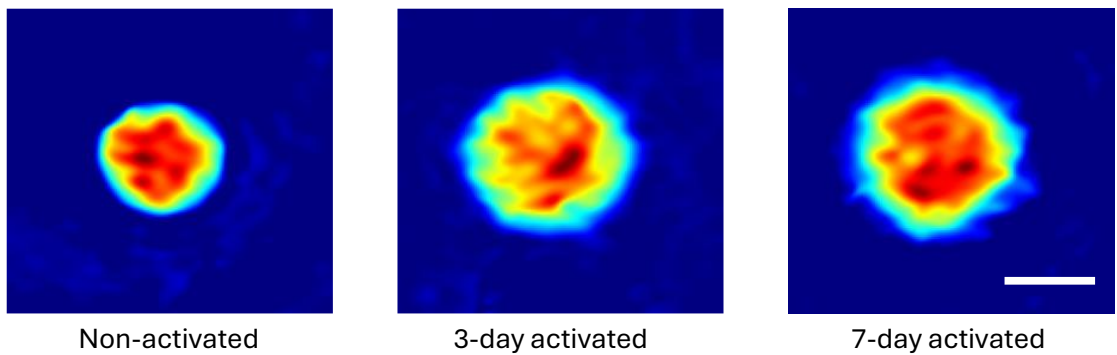

**Figure 5.** Representative cell types from the OPD images dataset. Notably, the activated cells at 3 and 7 days exhibit highly similar morphological characteristics, making visual distinction challenging. The white bar indicates 5 μm.

## 3.2 Classification Performance

To evaluate the effectiveness of our SparseAttnNet approach, we compared it against well-established CNN architectures: ResNet50 and MobileNetV2. These models were chosen as benchmarks due to their widespread use in medical image analysis and varying computational profiles. Although our method incorporates attention mechanisms conceptually like Vision Transformers (ViT), we did not include ViT in our quantitative comparisons due to its prohibitively high computational requirements for the high-resolution cell-microscopy images in our datasets. The quadratic complexity of self-attention in ViT with respect to the number of input tokens makes it impractical for deployment in resource-constrained environments, which is the scenario our method aims to address. Table 1 presents the classification performance across all datasets. The maximum possible of $k$ is the image size, which is 65,536 pixels for the brightfield microscopy, and 18,225 pixels for the rest of the datasets.

**Table 1.** Classification Performance

| Dataset | Model | Accuracy (%) | Precision (%) | Recall (%) | F1 Score | k Value | % Image |
|---|---|---|---|---|---|---|---|
| **Brightfield Microscopy** | Ours | 96.75 ± 0.23 | 96.77 ± 0.21 | 96.75 ± 0.23 | 0.9675 ± 0.0022 | 5529.6 ± 486.7 | 8.46 ± 0.74 |
| | Resnet50 | 97.36 ± 0.29 | 97.38 ± 0.29 | 97.36 ± 0.29 | 0.9736 ± 0.0029 | --- | 100 |
| | MobilenetV2 | 97.27 ± 0.43 | 97.33 ± 0.39 | 97.27 ± 0.43 | 0.9728 ± 0.0042 | --- | 100 |
| **Event** | Ours | 94.94 ± 0.45 | 94.96 ± 0.43 | 94.94 ± 0.45 | 0.9494 ± 0.0045 | 3040 ± 489.1 | 16.68 ± 2.68 |
| | Resnet50 | 94.43 ± 1.03 | 94.55 ± 1.04 | 94.43 ± 1.03 | 0.9443 ± 0.0103 | --- | 100 |
| | MobilenetV2 | 93.67 ± 1.30 | 93.70 ± 1.31 | 93.67 ± 1.30 | 0.9367 ± 0.0130 | --- | 100 |
| **OPD** | Ours | 94.05 ± 0.47 | 94.06 ± 0.47 | 94.05 ± 0.47 | 0.9403 ± 0.0047 | 2145.8 ± 570.4 | 11.76 ± 3.12 |
| | Resnet50 | 95.29 ± 0.38 | 95.35 ± 0.35 | 95.29 ± 0.38 | 0.9529 ± 0.0037 | --- | 100 |
| | MobilenetV2 | 94.54 ± 0.50 | 94.57 ± 0.54 | 94.54 ± 0.50 | 0.9453 ± 0.0051 | --- | 100 |
| **Raw Off-Axis Holograms** | Ours | 99.07 ± 0.36 | 99.08 ± 0.35 | 99.07 ± 0.36 | 0.9907 ± 0.0036 | 2143.2 ± 449.9 | 11.76 ± 2.46 |
| | Resnet50 | 99.60 ± 0.15 | 99.61 ± 0.15 | 99.60 ± 0.15 | 0.9960 ± 0.0015 | --- | 100 |
| | MobilenetV2 | 98.96 ± 0.22 | 98.96 ± 0.22 | 98.96 ± 0.22 | 0.9896 ± 0.0022 | --- | 100 |

Table 1 demonstrates that SparseAttnNet achieves competitive classification performance across all four datasets while processing only a fraction of the image pixels. For holographic images, our model attains 99.07% accuracy, which is comparable to the state-of-the-art ResNet50 (99.60%) and slightly better than MobileNetV2 (98.96%), while utilizing only 11.76% of the image pixels. The value of $k$ for this dataset converged to approximately 2143 pixels, indicating that the model can effectively extract discriminative features from a sparse representation of the original image.

For event camera images, SparseAttnNet slightly outperforms ResNet50 with 94.94% accuracy compared to 94.43%, and significantly surpasses MobileNetV2 (93.67%). This superior performance is achieved while processing just 16.68% of the pixels, demonstrating our model ability to effectively handle the inherently sparse nature of event camera data. The higher percentage of selected pixels compared to other datasets aligns with the sparse representation already present in event camera outputs, where only pixels detecting motion are recorded.

In the case of OPD images, SparseAttnNet achieves 94.05% accuracy, which is slightly lower than ResNet50 (95.29%) but comparable to MobileNetV2 (94.54%). This slight performance gap can be attributed to the subtle morphological differences between activated T-cells at different time points, which might require more context than what is captured by the sparse sampling. Nevertheless, our model maintains competitive performance while processing only 11.76% of pixels.

For the more challenging database of conventional brightfield microscopy of stained WBCs, having significant class imbalance, SparseAttnNet achieves 96.75% accuracy, which is within 1% of ResNet50 (97.36%) and MobileNetV2 (97.27%). This dataset required a higher $k$ value (5529.6 pixels on average), yet this still represents only 8.43% of the total image pixels. The higher pixel requirement likely reflects the complexity of distinguishing between various WBC types, particularly for underrepresented classes such as basophils.

Across all datasets, the average standard deviation of accuracy for SparseAttnNet is comparable to ResNet50 and MobileNetV2, indicating similar stability in classification performance. The precision, recall, and F1 scores closely mirror the accuracy trends, confirming balanced performance across different evaluation metrics.

A notable observation is the adaptive nature of our approach in determining the optimal number of $k$ pixels for each dataset. The values of $k$ vary significantly based on the image complexity and information density: from approximately 2145 pixels for holographic and OPD images to 5529 pixels for the brightfield microscopy dataset with its complex cellular morphology. This adaptability is a key strength of SparseAttnNet, allowing it to allocate computational resources based on the inherent complexity of the classification task.

## 3.3 Visualization of Attention Maps

To gain insights into how SparseAttnNet focuses on specific regions within cell images, we visualized the distilled coarse attention, the $k$ pixels selector, and the fine attention across our datasets. The attention maps consistently highlight biologically significant regions such as cell membranes, nuclear boundaries, and cytoplasmic structures.

In off-axis holographic images, as can be seen in Fig. 6, the attention focuses on three-dimensional cellular morphology features that are characteristic of different T-cell states. The interference patterns captured in these images provide additional structural information that the model leverages effectively, as evidenced by attention hotspots on phase-dense regions. We also added the paired OPD image, to show that the fine attention focus on the fringes that are located on the cell area.

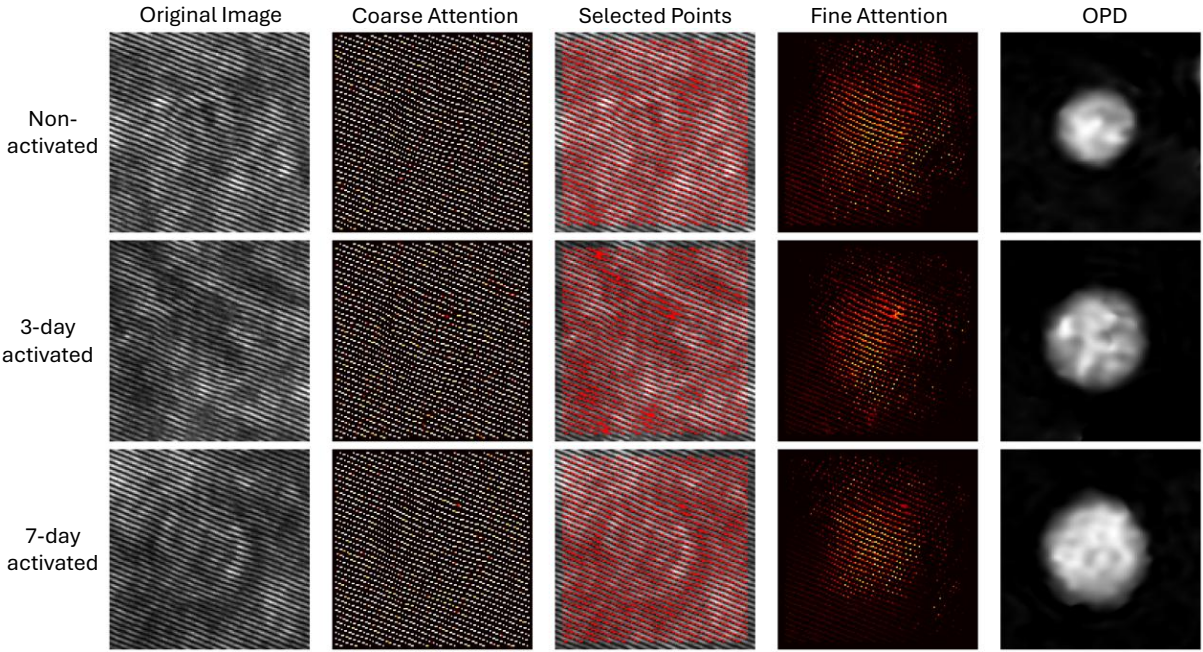

**Figure 6.** Visualization of the hierarchical attention mechanism using off-axis holographic images. The attention maps reveal how our model focuses on areas where interference fringes interact with cell boundaries, containing crucial phase information despite the complex background patterns. Note how the fine attention map specifically targets regions of fringe modulation at the cell periphery while ignoring the uniform carrier fringe patterns in the background.

In event camera images, as can be seen in Fig. 7, despite the sparse nature of the data, the attention mechanism correctly identifies motion-sensitive regions that differentiate cancer cell types. For SW480 cells (primary tumor), the attention module concentrates on the more regular cellular boundaries, while for SW620 cells (metastatic), the attention module shifts to the irregular membrane protrusions characteristic of invasive phenotypes.

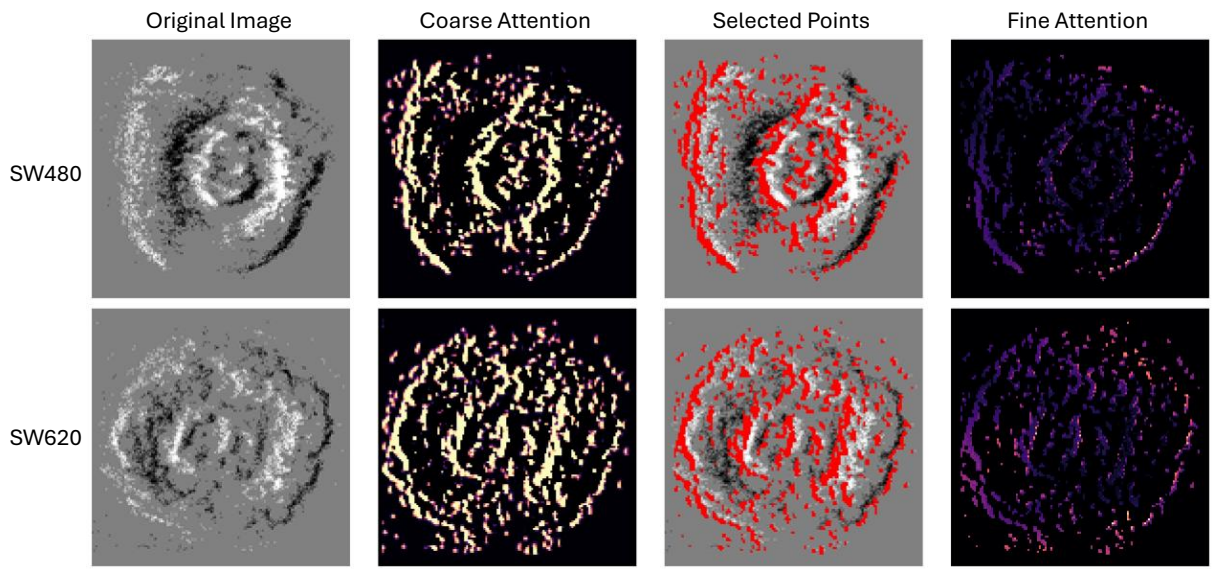

**Figure 7.** Visualization of the hierarchical attention mechanism using event camera images. Despite the inherently sparse nature of event data, our model successfully identifies morphological differences between SW480 primary tumor cells (more regular boundaries) and SW620 metastatic cells (irregular membrane protrusions and invasive phenotype). The attention patterns highlight motion-sensitive regions that contain discriminative features for classification.

In brightfield microscopy images, as can be seen in Fig. 8, where the attention mechanisms precisely target the class-defining nuclear morphology: the large, round nucleus of lymphocytes, the distinctive bi-lobed nucleus of eosinophils, and the multi-lobed nucleus of neutrophils. The model also refers to cytoplasmic granularity patterns that are diagnostic hallmarks for differentiating granulocyte subtypes.

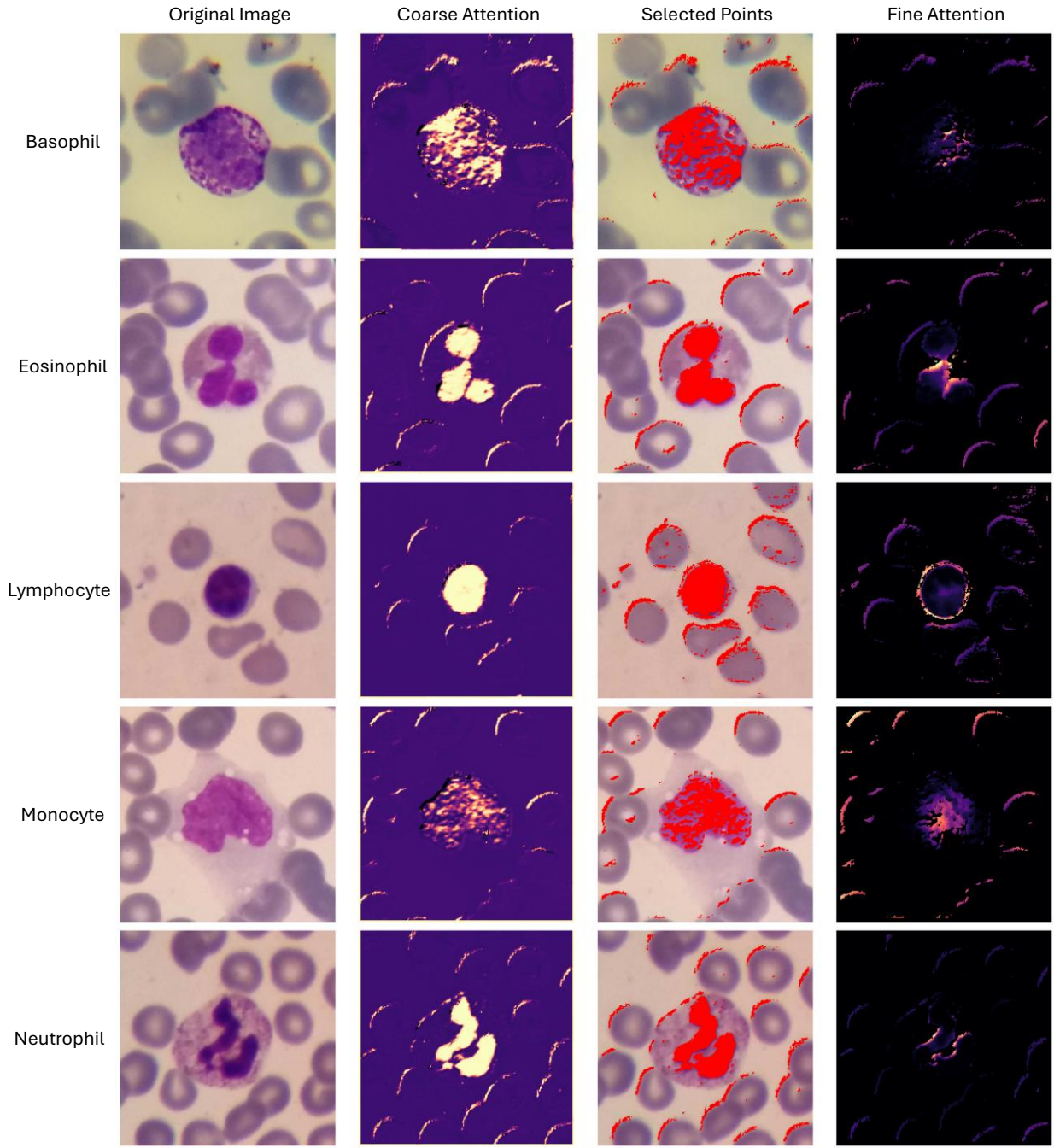

**Figure 8.** Visualization of the hierarchical attention mechanism using brightfield microscopy images. The attention maps demonstrate cell-specific focus patterns that align with hematological diagnostics: lymphocytes (large round nuclei), monocytes (kidney-shaped nuclei), neutrophils (multi-lobed nuclei), basophils (coarse granules), and eosinophils (bi-lobed nuclei with distinctive granules). These visualizations confirm that our model identifies the same diagnostically relevant features used by hematologists.

In the OPD T-cell images, as can be seen in Fig. 9, the density of the pixels correlates with the complexity of cellular structures: higher densities appear in areas with more intricate morphological features, such as the nuclear boundaries in WBCs and the activation-induced membrane projections in T-cells (Cristofanilli et al., 2004; Santana and Esquivel-Guadarrama, 2006).

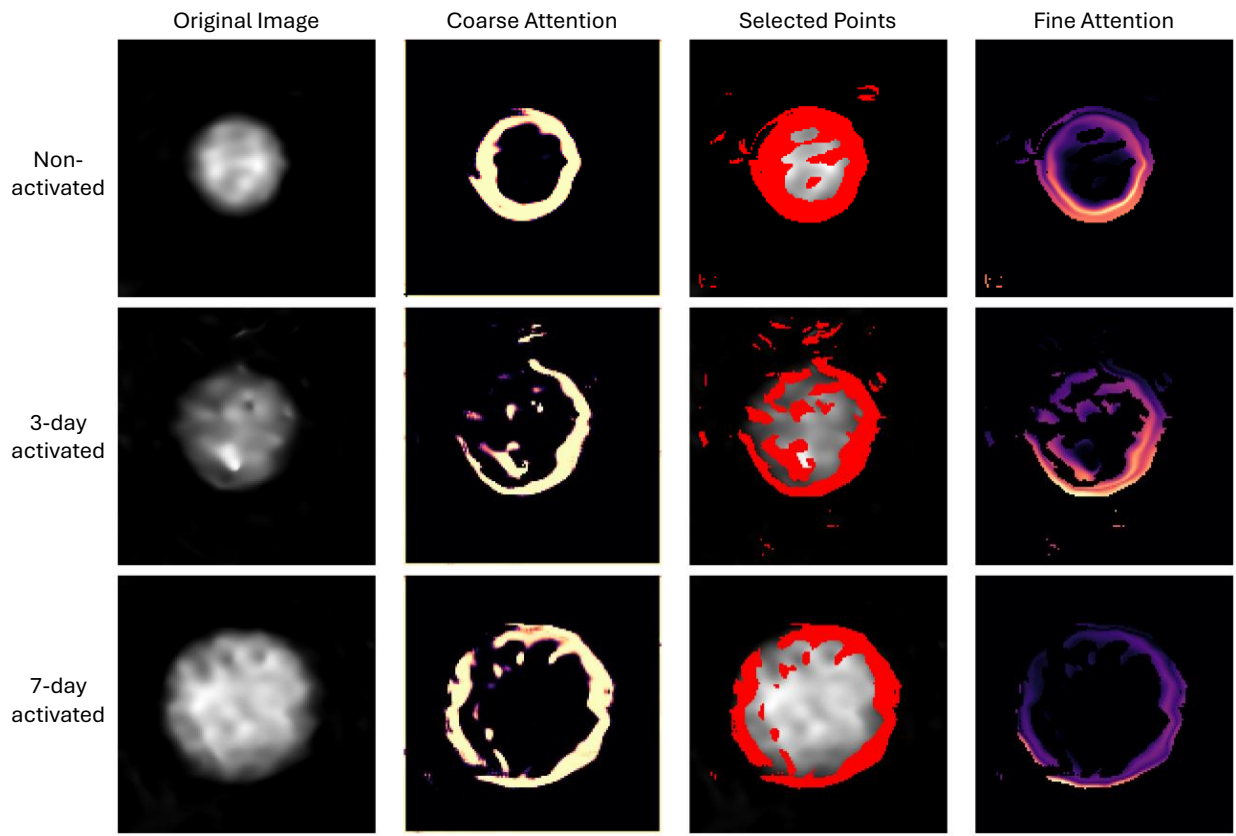

**Figure 9.** Visualization of the hierarchical attention mechanism using OPD images. Our model consistently focuses on the cell periphery and subcellular structures rather than central regions, aligning with biological insights about T-cell activation which manifests through peripheral cytoskeletal rearrangements and membrane dynamics. The attention maps show higher concentration at cell boundaries where activation-induced structural remodeling generates distinctive optical signatures.

Across all datasets, the attention mechanisms effectively suppress background regions, directing computational resources only to the informative parts of the images. This selective focus is crucial for computational efficiency, as background areas typically constitute up to approximately 90% of microscopy fields of view but contribute minimally to classification decisions.

These visualizations provide strong evidence that SparseAttnNet enhances model explainability by focusing computational resources on biologically meaningful features. The model's selective attention mechanism not only improves efficiency by processing only ~15% of the pixels but also increases interpretability by highlighting diagnostically relevant cellular structures. This explainability aspect is crucial for building trust in AI-assisted microscopy analysis, particularly in clinical applications where understanding the basis for classification decisions is essential for adoption by healthcare professionals.

## 3.4 Computational Efficiency

To evaluate the computational advantages of SparseAttnNet, we conducted a comprehensive analysis comparing its efficiency against established CNN architectures (ResNet50 and MobileNetV2) in terms of parameters, FLOPs, inference time, and throughput on both CPU and GPU hardware for image size of 135×135. We excluded Vision Transformers (ViT) from our quantitative efficiency comparison due to their prohibitively high computational demands for these image resolutions. A standard ViT-Base/16 model for 135×135 images would require approximately 86 million parameters and 8.4 billion FLOPs, making it impractical for resource-constrained environments and real-time applications like flow cytometry. To evaluate the computational advantages of SparseAttnNet, we conducted a comprehensive analysis comparing its efficiency against established CNN architectures (ResNet50 and MobileNetV2) in terms of parameters, FLOPs, inference time, and throughput on both CPU and GPU hardware for image size of 135×135. The results are summarized in Table 2.

**Table 2.** Computational efficiency comparison between SparseAttnNet and established CNN architectures. The table reports model size (parameters in millions), computational complexity (FLOPs in billions), and inference performance on both CPU and GPU hardware (inference time in milliseconds and throughput in images per second). SparseAttnNet achieves substantial efficiency gains across all metrics while maintaining competitive classification accuracy, demonstrating the effectiveness of our sparse processing approach.

| Model | Parameters (M) | FLOPs (G) | CPU Time (ms) | CPU Throughput (image/s) | GPU Time (ms) | GPU Throughput (image/s) |
|---|---|---|---|---|---|---|
| Ours | 0.05 | 0.04 | 419.87 | 152.43 | 5.51 | 11610.76 |
| ResNet50 | 23.51 | 1.69 | 12429.09 | 5.15 | 27.01 | 2369.76 |
| MobileNetV2 | 2.23 | 0.14 | 2673.61 | 23.94 | 10.04 | 6373.00 |

SparseAttnNet demonstrates remarkable efficiency with only 0.05 M parameters, which is approximately 50× fewer than MobileNetV2 (2.23M) and 470× fewer than ResNet50 (23.51 M). This dramatic reduction in model size makes SparseAttnNet particularly suitable for deployment in memory-constrained environments such as point-of-care diagnostic devices.

In terms of computational complexity, our model requires only 0.04G FLOPs, which is 3.5× lower than MobileNetV2 (0.14G) and 42× lower than ResNet50 (1.69G). This reduction is directly attributable to our sparse processing approach that focuses computational resources exclusively on the most informative regions of the image.

The practical implications of these efficiency gains are evident in the inference measurements. On CPU, SparseAttnNet achieves an inference time of 419.87 ms per image, significantly faster than MobileNetV2 (2673.61 ms) and ResNet50 (12429.09 ms). This translates to a throughput of

152.43 images per second, which is 6.4× higher than MobileNetV2 (23.94) and 29.6× higher than ResNet50 (5.15).

The efficiency gains are also evident on GPU hardware, though it should be noted that the comparison on this platform understates SparseAttnNet's true potential. While ResNet50 and MobileNetV2 benefit from years of GPU optimization and hardware-specific acceleration through highly optimized CUDA implementations, our current SparseAttnNet implementation has not yet undergone similar optimization efforts. Despite this disadvantage, SparseAttnNet still achieves an inference time of just 5.51 ms per image, compared to 10.04 ms for MobileNetV2 and 27.01 ms for ResNet50. The resulting throughput of 11610.76 images per second is 1.8× higher than MobileNetV2 (6373.00) and 4.9× higher than ResNet50 (2369.76). These results suggest that with dedicated optimization for GPU execution, the performance gap would likely widen even further in favor of our approach.

These performance metrics highlight how SparseAttnNet approach of processing approximately only the most informative 15% of image pixels translates into substantial real-world efficiency gains. The dynamic selection of sparse pixels eliminates redundant computation on uninformative background regions without sacrificing classification accuracy. This efficiency makes SparseAttnNet particularly well-suited for high-throughput applications such as automated imaging flow cytometry, where speed and resource utilization are critical considerations.

## 4. Discussion

Our approach maintains effectiveness even when discriminative features are distributed throughout the image, as the dynamic selection of $k$ pixels can adapt to capture these distributed patterns. While SparseAttnNet demonstrates efficiency and accuracy for cell classification, several important limitations warrant discussion. The key limitation arises when the optimal number of pixels $k$ approaches the total number of pixels in the image. In such scenarios, our method loses its computational advantage, as processing nearly all pixels negates the efficiency benefits of sparse attention. This situation could occur with highly complex cell types where almost every pixel contains relevant information, or with imaging modalities that produce inherently information-dense representations where feature redundancy is minimal. When $k$ is small (e.g., 15% of total pixels as in our experiments), the efficiency gains are substantial. However, if a particular classification task required $k$ to approach 80-90% of total pixels, the overhead of the attention mechanisms might actually increase computational cost compared to standard convolutional approaches that process all pixels directly without the selection step.

In addition, the dynamic $k$ selection mechanism assumes relatively consistent information density across the dataset. When dealing with heterogeneous data including both simple cells with localized features and complex cells with distributed features, a single global $k$ value may not be optimal for all instances. Some images might require higher pixel density to maintain accuracy.

These limitations highlight an important consideration for deploying SparseAttnNet in practice. This approach is most valuable for applications where relevant information is concentrated in a subset of the image or where the redundancy in the data allows for effective sparse sampling without significant loss of information.

To address these limitations in case they arise for specific datasets, future work could include an adaptive per-image $k$ selection based on individual image complexity, potentially with a cost-aware mechanism that balances accuracy and computational efficiency. In addition, hybrid dense-sparse processing can dynamically switch between sparse and dense processing based on detected information density. Information density estimation can be performed as a preprocessing step to determine whether sparse processing would be beneficial for a particular image or dataset. Finally, hardware-aware implementations can be optimized for specific computing platforms to maximize efficiency gains on GPUs and other specialized hardware.

## 5. Conclusion

We have presented SparseAttnNet, a novel hierarchical attention-driven framework for efficient cell classification that dynamically selects and processes only the most informative pixels from microscopy images. Our approach fundamentally reimagines how deep learning systems can interact with image data by shifting away from exhaustive pixel processing toward a selective, information-driven paradigm. Our method includes a dynamic pixel selection mechanism that adaptively determines the optimal number of pixels in the image to process, automatically balancing between computational efficiency and classification performance. It is based on a hierarchical attention architecture that progressively refines focus from coarse to fine attention, effectively distilling information across levels. Our integrated loss formulation combines focal classification, contrastive representation learning, and attention distillation to guide the model toward identifying biologically relevant features. When applied to diverse cell optical microscopy classification tasks using conventional brightfield-microscopy stained cells, sparse event-camera microscopy of unstained cells, and highly informative digital holographic microscopy of unstained cells, SparseAttnNet achieved classification accuracy comparable to or exceeding that of conventional CNN architectures while requiring only a fraction of the computational resources. Processing approximately 15% of the pixels reduced parameter count by up to 470× and FLOPs by up to 42× compared to ResNet50, demonstrating that intelligent feature selection can dramatically improve computational efficiency without sacrificing performance. Beyond efficiency, our approach enhances explainability by highlighting the specific cellular regions that drive classification decisions. The sparse attention maps show a strong alignment with biologically meaningful structures, providing transparency often lacking in conventional 'black-box' deep learning models. For example, in the OPD images, the model consistently focuses on the inner margins of the cell rather than the central region when distinguishing between activated and non-activated T cells. This behavior aligns with known biological processes, as activation induces peripheral structural remodeling such as actin reorganization and membrane dynamics that primarily affect the inner cell boundary. These changes result in localized variations in the cell refractive index and its thickness, generating distinct optical signatures in the OPD maps. As a result, the inner margins offer more informative and discriminative features than the relatively uniform center, making them key regions for accurate classification.

SparseAttnNet opens new possibilities for deploying sophisticated cell analysis in resource-constrained environments, enabling high-throughput processing on modest hardware, and paving the way for more interpretable AI systems in clinical cell diagnosis and biomedical research. This selective processing approach represents a promising direction for medical image analysis that aligns computational focus with biologically relevant information, a principle that may extend well beyond cell classification to broader domains of biomedical image analysis.

## 6. Data Availability

The data of the holographic images, OPD images, event camera images are available upon request. The WBC dataset has open-access and is available as cited in (Nickparvar, 2022).

## Funding

Supported by Israel Innovation Authority.

## Keywords

Sparse attention, Cell classification, Knowledge distillation, Computational efficiency, Hierarchical attention, Microscopy image analysis, Flow cytometry.